\documentclass[twocolumn,prc,showpacs,preprintnumbers,amsmath,amssymb,floatfix]{revtex4}
\usepackage{amsmath}
\usepackage{graphicx}
\usepackage{dcolumn}
\usepackage{bm}
\usepackage{ulem} 
\usepackage[usenames]{color}
\usepackage{epstopdf}
\usepackage{epsfig}
\usepackage{float}
\usepackage{subfigure}

\begin{document}

\title{Nucleon radius effects on neutron stars in quark mean field model}
\author {Zhen-Yu Zhu}
\author {Ang Li}
\email{liang@xmu.edu.cn}
\affiliation{Department of Astronomy, Xiamen University, Xiamen 361005, China} 

\date{\today}

\begin{abstract}

We study the effects of free space nucleon radius on nuclear matter and neutron stars within the framework of quark mean field model. The nucleon radius is treated self-consistently with this model, where quark confinement is adjusted to fit different values of nucleon radius. Corrections due to center-of-mass motion, quark-pion coupling, and one gluon exchange are included to obtain the nucleon mass in vacuum. The meson coupling constants that describe the behavior of many-body nucleonic system are newly-constructed by reproducing the empirical saturation properties of nuclear matter, including the recent determinations of symmetry energy parameters. Our results show that the nucleon radius in free space have negligible effects on nuclear matter equation of state and neutron star mass-radius relations, which is different from the conclusion drawn in previous studies. We further explore that the sensitivity of star radius on the nucleon radius found in earlier publications~\cite{proton,nuclradius} is actually from the symmetry energy and its slope.

\end{abstract}


\maketitle
\section{Introduction}

The equation of state (EoS) of cold dense matter has attracted much attention of astronomers and nuclear physicists, because it governs the structures of neutron stars (NSs) and is still uncertain due to the poorly-known nature of strong interaction. The most controversial high density part of the EoS might be probed through the ultra-dense matter present in NSs' cores and terrestrial heavy-ion collision experiments~\cite{nmexperiment}. The information of EoS of a NS is encoded in the mass-radius relation that can be extracted from the observational data. Therefore, the study on mass-radius relations of NSs can help us determine the EoS of cold ultra-dense matter. Up to now, masses of more than three dozen NSs have been measured relatively precisely, see e.g., ~\cite{lattimermass,ozelmassradius}, but the estimation of NS radii from observational data is highly uncertain, see discussions in e.g.,~\cite{ozelmassradius}. Simultaneous measurements of both mass and radius for one NS is even more difficult. 

A recent work in Ref.~\cite{proton} connected the NS radius to the properties of nucleons, and concluded that the free nucleon radius $r_N$ could significantly affect the NS radius $R$. The authors extended later their work in Ref.~\cite{nuclradius} to slowly rotating NSs, taking into account more constrains on nuclear matter properties beyond the saturation density $\rho_0$. The sensitivity of $R$ on $r_N$ was similar. In the present paper we aim to reexamine this dependence within an alternative framework. Below we explain our concerns and motivations.

In the preceding papers~\cite{proton,nuclradius}, in order to explain the proton radius puzzle (see the most recent discussions in e.g.,~\cite{ppuzzle1,ppuzzle2}), the authors suggested that protons could have a distribution of radii rather than a fixed size. They varied $r_N$ from their fitted value $0.833$ fm (later updated to $0.864$~\cite{nuclradius1}) by $\pm 20\%$, and made use of a phenomenological density-dependent form of $r_N (\rho) = r_N/[1+\beta (\rho/\rho_0)^2]^2$ in relativistic mean field (RMF) model for nucleon radius in medium. The free parameter $\beta$ is chosen to accord with the results from the quark-meson coupling (QMC) model using a constant bag parameter ($B=\rm{const}$) in Ref.~\cite{radiusform}. 

However, as found previously in Refs.~\cite{excludevolume,mbag1,mbag2,mbag3,mbag4}, the medium modification of the bag constant might play an important role in low- and medium-energy nuclear physics. In particular, in the two cases of bag parameter, i.e., constant or in-medium changed ($B=B^*$), the density-dependence behaviour of $r_N (\rho)$ is opposite. Therefore it could be very interesting to study the above-mentioned $r_N$-vs-$R$ dependence using different $r_N (\rho)$ result. It may reveal important insights of quark structure effects in a nucleon, which is crucial for understanding better the short-range properties of nucleon-nucleon interaction (usually simplified as excluded volume effect in phenomenological models like in Ref.~\cite{excludevolume}). Moreover, in the previous RMF study, the protons and neutrons are treated as pointlike particle, so the excluded volume effect and the density-dependence of the nucleon radius, $r_N (\rho)$, can not be incorporated consistently in the model, and were adapted from the results of another QMC calculation.

Above all, for the purpose of the present work, we employ the quark mean field (QMF) model (e.g., \cite{tokiqmf,shenqmf,huqmf,hunsqmf,xing}) where constituent quarks ($m_q = 300$ MeV) are confined with a potential in the harmonic oscillator form. The quark potential has also been previously employed in e.g., Refs.~\cite{panda1,panda0,panda2,panda3}. The free space radius of nucleons (three values are chosen around the CODATA 2014 value~\cite{codata}), as well as their mass ($m_N = 939$ MeV), are our input for determining the quark potential parameters. Its density dependence $r_N (\rho)$ are consistently deduced from solving the Dirac equation for a nucleon in medium. Then nucleons interact with each other through $\sigma, \omega, \rho$ meson fields, with various meson coupling constants newly fitted from the empirical values of ($\rho_0, E/A, J, K, L, M_N^*/M_N$) at nuclear matter saturation point. The study of nuclear matter and NSs can be followed, and the influence of the employed $r_N$ on nuclear matter EoS and NS properties can then be addressed consistently within QMF.

In Sec. II we briefly introduce the QMF model and the fitting of model parameters; In Sec. III, results and discussions are displayed. Finally, we will summarize our work and conclude in Sec. IV.

----------------------------------------------------------------
\section{Formalism}

\subsection{QMF model}
In the following, we will give a brief introduction to the QMF model. This model starts with a flavor independent potential $U(r)$ confining the constituent quarks inside the nucleon. The potential has a harmonic oscillator form
\begin{eqnarray}
U(r)=\frac{1}{2}(1+\gamma^0)(ar^2+V_0),
\end{eqnarray}
with the parameters $a$ and $V_0$ to be determined. The Dirac equation of the confined quarks is written as
\begin{eqnarray}
[\gamma^{0}(\epsilon_{q}-g_{\omega q}\omega-\tau_{3q}g_{\rho q}\rho)-\vec{\gamma}\cdot\vec{p}  \nonumber \\
-(m_{q}-g_{\sigma q}\sigma)-U(r)]\psi_{q}(\vec{r})=0.
\end{eqnarray}
Hereafter $\psi_{q}(\vec{r})$ is the quark field, $\sigma$, $\omega$, and $\rho$ are the classical meson fields. $g_{\sigma q}$, $g_{\omega q}$, and $g_{\rho q}$ are the coupling constants of $\sigma, ~\omega$, and $\rho$ mesons with quarks, respectively. $\tau_{3q}$ is the third component of isospin matrix. This equation can be solved exactly and its ground state solution for energy is
\begin{eqnarray}
(\mathop{\epsilon'_q-m'_q})\sqrt{\frac{\lambda_q}{a}}=3,
\end{eqnarray}
where $\lambda_q=\epsilon_q^\ast+m_q^\ast,\ \mathop{\epsilon'_q}=\epsilon_q^\ast-V_0/2,\ \mathop{m'_q}=m_q^\ast+V_0/2$. The effective single quark energy is given by $\epsilon_q^*=\epsilon_{q}-g_{q\omega}\omega-\tau_{3q}g_{q\rho}\rho$ and the effective quark mass by $m_q^\ast = m_q-g_{\sigma q}\sigma$ with the quark mass $m_q$ = 300 MeV.
The solution for wave function is
\begin{eqnarray}
\psi_{q}(r,\theta,\phi)=\frac{1}{r} \left(
\begin{array}{c}
F(r)Y_{1/2 m}^0(\theta,\phi) \\
iG(r)Y_{1/2 m}^1(\theta,\phi)
\end{array}
\right),
\end{eqnarray}
where
\begin{eqnarray}
F(r)=\mathcal{N}\left(\frac{r}{r_{0}}\right)\exp(-r^2/2r_0^2), \nonumber \\
G(r)=-\frac{\mathcal{N}}{\lambda_qr_0}\left(\frac{r}{r_{0}}\right)^2\exp(-r^2/2r_0^2), \nonumber
\end{eqnarray}
\begin{eqnarray}
r_0=(a\lambda_q)^{-1/4},\ \ \ \mathcal{N}^2=\frac{8\lambda_q}{\sqrt{\pi}r_0}\frac{1}{3\mathop{\epsilon'_q}+\mathop{m'_q}}.\nonumber
\end{eqnarray}

\begin{table}\label{tab1}
\begin{center}
\caption{Values of the free nucleon radius $r_N$ used in this work and the corresponding parameter ($a$ and $V_0$) for quark potential in Eq. (1). The intermediate value $0.87$ fm is from~\cite{codata}. The free nucleon mass is fixed to $M_N = 939$ MeV.}
\vskip 0.3cm
\begin{tabular}{c|cc}\hline
$r_N$ [fm] & $a$ [fm$^{-3}$] & $V_0$ [MeV] \\ \hline
0.80 & 0.735186 & -71.565596 \\
0.87 & 0.534296 & -62.257187 \\
1.00  & 0.312494 & -48.389200 \\ \hline
\end{tabular}
\end{center}
\end{table}

The zeroth-order energy of the nucleon core $E_N^0=\sum_q\epsilon_q^\ast$ can be obtained by solving Eq. (2). The contribution of center-of-mass correction $\epsilon_{c.m.}$, pionic correction $\delta M_N^\pi$ and gluonic correction $(\Delta E_N)_g$ are also taken into account following Refs. \cite{xing,panda1}. 

For the center-of-mass correction, the energy contribution can be written as:
\begin{eqnarray}
\epsilon_{c.m.}=\frac{\mathop{77\epsilon'_q + 31m'_q}}{3(\mathop{3\epsilon'_q + m'_q})^2r_0^2}.
\end{eqnarray}

For pionic correction,
\begin{eqnarray}
\delta M_N^\pi=-\frac{171}{25}I_\pi f^2_{NN\pi},
\end{eqnarray}
where
\begin{eqnarray}
I_\pi=\frac{1}{\pi m_\pi^2}\int_0^{\infty}dk\frac{k^4u^2(k)}{k^2+m_\pi^2}, \nonumber
\end{eqnarray}
\begin{eqnarray}
u(k)=\left[1-\frac{3}{2}\frac{k^2}{\lambda_q(\mathop{5\epsilon'_q + 7m'_q})}\right]\exp(-\frac{1}{4}r_0^2k^2), \nonumber
\end{eqnarray}
and
\begin{eqnarray}
f_{NN\pi}=\frac{25\epsilon'_q + 35m'_q}{27\epsilon'_q + 9m'_q} \frac{m_\pi}{4\sqrt{\pi}f_\pi}. \nonumber
\end{eqnarray}
The constants $m_\pi=140$ MeV and $f_\pi=93$ MeV are the mass of $\pi$ meson and the phenomenological pion decay constant, respectively. 

For gluonic correction,
\begin{eqnarray}
(\Delta E_N)_g=-\alpha_c\left(\frac{256}{3\sqrt{\pi}}\frac{1}{R_{uu}^3}\frac{1}{(\mathop{3\epsilon'_q + m'_q})^2}\right),
\end{eqnarray}
where
\begin{eqnarray}
R_{uu}^2=\frac{6}{\mathop{\epsilon'^2_q - m'^2_q}} \nonumber
\end{eqnarray}
and $\alpha_c=0.58$ is a constant.

With these corrections on energy, we can then determine the mass of nucleon:
\begin{eqnarray}
M^\ast_N=E^{0}_N-\epsilon_{c.m.}+\delta M_N^\pi+(\Delta E_N)_g.
\end{eqnarray}

The nucleon radius in the QMF theory is written as
\begin{eqnarray}
\langle r_N^2\rangle = \frac{\mathop{11\epsilon'_q + m'_q}}{\mathop{(3\epsilon'_q + m'_q)(\epsilon'^2_q-m'^2_q)}}.
\end{eqnarray}

As mentioned in the introduction, the potential parameters ($a$ and $V_0$) in Eq.~(1) are obtained from reproducing the nucleon mass and radius $(m_N, r_N)$ in free space. To study the $r_N$ effect, we vary it from the intermediate value $0.87$ fm~\cite{codata} by around $10 \%$ according to our model capability: $r_N$ = 0.80 fm, 0.87 fm, 1.00 fm. It covers both two latest experimental analyses of the $rms-$radius of the proton charge distribution: 0.879 $\pm$ 0.009 fm~\cite{exp1} from electron-proton scattering and 0.8409 $\pm$ 0.0004 fm~\cite{exp2} from the Lamb shift measurement in muonic hydrogen. A latest estimation of $\sim 0.81$ fm can also be described~\cite{rn18}. The employed $(m_N, r_N)$ input and the corresponding results of potential parameters ($a$ and $V_0$) are shown in Table 1.

\subsection{Meson-coupling parameters}

For the study of infinite nuclear matter, from this step nucleons are treated as point-like particles and interact through exchange of $\sigma,~\omega,~\rho$ mesons. The Lagrangian is written as (e.g.,\cite{hunsqmf,xing}):
\begin{eqnarray}
\mathcal{L}& = & \overline{\psi}\left(i\gamma_\mu \partial^\mu - M_N^\ast - g_{\omega N}\omega\gamma^0 - g_{\rho N}\rho\tau_{3}\gamma^0\right)\psi \nonumber \\ 
& & -\frac{1}{2}(\nabla\sigma)^2 - \frac{1}{2}m_\sigma^2 \sigma^2 - \frac{1}{4}g_3\sigma^4 + \frac{1}{2}(\nabla\rho)^2 + \frac{1}{2}m_\rho^2\rho^2 \nonumber \\ 
& & + \frac{1}{2}(\nabla\omega)^2 + \frac{1}{2}m_\omega^2\omega^2 + \frac{1}{4}c_3\omega^4 + \frac{1}{2}g_{\rho N}^2\rho^2 \Lambda_v g_{\omega N}^2\omega^2 \nonumber
\end{eqnarray}
where $g_{\omega N}$ and $g_{\rho N}$ are the nucleon coupling constants for $\omega$ and $\rho$ mesons. The quark counting rule gives $g_{\omega N}=3g_{\omega q}$ and $g_{\rho N}=g_{\rho q}$~\cite{shen}. The calculation of confined quarks in the previous section gives the relation of effective nucleon mass $M_N^*$ as a function of $\sigma$ field, which defines the $\sigma$ coupling with nucleons (depending on the parameter $g_{\sigma q}$). $m_{\sigma},~m_{\omega}$, and $m_{\rho}$ are the meson masses.
The last term of the Lagrangian, is the cross coupling from $\omega$ meson and $\rho$ meson~\cite{horowitz}. It is introduced in this work to give a reasonable slope of symmetry energy~\cite{Lireview,panda12}. 

\begin{table}\label{tab2}
\begin{center}
\caption{Saturation properties used in this study for the fitting of new sets of nucleon-meson coupling parameters: The saturation density $\rho_0$ (in fm$^{-3}$) and the corresponding values at saturation point for the binding energy $E/A$ (in MeV), the incompressibility $K$ (in MeV), the symmetry energy $J$ (in MeV), the symmetry energy slope $L$ (in MeV) and the ratio between the effective mass and free nucleon mass $M_N^\ast/M_N$.}
\vskip 0.3cm
\begin{tabular}{c|ccccc}\hline
$\rho_0$ & $E/A$ & $K$ & $J$ & $L$ & $M_N^\ast/M_N$ \\ 
$[{\rm fm}^{-3}]$ & [MeV] & [MeV] & [MeV] & [MeV] & / \\\hline
0.16 & -16 & 220/260 & 28/31/34 & 40/60/80 & 0.74 \\\hline
\end{tabular}
\end{center}
\end{table}

There are six parameters ($g_{\sigma q}, g_{\omega q}, g_{\rho q}, g_3, c_3, \Lambda_v$) in this Lagrangian and they will be determined by fitting the saturation density $\rho_0$ and the corresponding values at saturation point for the binding energy $E/A$, the incompressibility $K$, the symmetry energy $J$, the symmetry energy slope $L$ and the effective mass $M_N^\ast$. Those employed values are collected in Table 2. In particular, we use the most preferred values for $(K, J, L)$ as recently suggested in Ref.~\cite{baoan,steiner}, namely $K = 240 \pm 20$ MeV, $J = 31.6 \pm 2.66$ MeV, $L = 58.9 \pm 16$ MeV. A recent fitting from finite nuclei data in the same model had $K = 328$ MeV~\cite{xing}, and we choose this case as well for comparison. 
For each $r_N$ value in Table 1, we first determine the potential parameters ($a$ and $V_0$) from reproducing $(m_N, r_N)$, then determine many-body parameters from reproducing the saturation properties of nuclear matter ($\rho_0, E/A, J, K, L, M_N^*/M_N$). Finally, 81 new QMF parameter sets are newly fitted for studying nuclear matter and compact stars from the quark level. The details of the ($g_{\sigma q}, g_{\omega q}, g_{\rho q}, g_3, c_3, \Lambda_v$) results, comparing to other versions of QMF theory, will be published in a separate paper~\cite{zhu}.

The equations of motion for mesons can be obtained by variation of the Lagrangian,
\begin{eqnarray}
m_\sigma^2\sigma + g_3\sigma^3 & = & (-\frac{\partial M^*_N}{\partial \sigma})\rho_S, \\
m_\omega^{\ast2}\omega + c_3\omega^3 & = & g_{\omega N}\rho_N, \\
m_\rho^{\ast2}\rho & = & g_{\rho N}\rho_3,
\end{eqnarray}
where
\begin{eqnarray}
\rho_S & = & \frac{1}{\pi^2}\sum_{i=n,p}\int_0^{p_F^i}dpp_i^2\frac{M^\ast_N}{\sqrt{M^{\ast2}_N+p_i^2}} \nonumber \\
 & = & \frac{M_N^\ast}{2\pi^2} \left(p_F^i E_F^i - M_N^{\ast2}\ln\bigg|\frac{p_F^i+E_F^i}{M_N^\ast}\bigg|\right), \nonumber
\end{eqnarray}
\begin{eqnarray}
E_F^i=\sqrt{M_N^{\ast2}+(p_F^i)^2} \nonumber,
\end{eqnarray}
\begin{eqnarray}
m_\omega^{\ast2}=m_\omega^2+\Lambda_vg_{\omega N}^2g_{\rho N}^2\rho^2, \\ \nonumber m_\rho^{\ast2}=m_\rho^2+\Lambda_vg_{\rho N}^2g_{\omega N}^2\omega^2. \nonumber
\end{eqnarray}
$p_F^n~(p_F^p$) is the Fermi momentum for neutron~(proton), $\rho_N = \rho_p+\rho_n $ and $\rho_3 = \rho_p-\rho_n$ that equals $0$ in symmetric nuclear matter. 

With known meson fields from Eqs.~(10)-(12), the Hamiltonian
\begin{eqnarray}
\mathcal{H} & = & \frac{1}{\pi^2}\sum_{i=n,p}\int_0^{p_F^i}\sqrt{p^2+M_N^{\ast2}}p^2dp + g_{\omega N}\omega\rho_N + g_{\rho N}\rho\rho_3 \nonumber \\
& & + \frac{1}{2}m_\sigma^2\sigma^2 + \frac{1}{4}g_3\sigma^4 - \frac{1}{2}m_\omega^2\omega^2-\frac{1}{4}c_3\omega^2 - \frac{1}{2}m_\rho^2\rho^2 \nonumber \\
& & - \frac{1}{2}\Lambda_vg_{\rho N}^2g_{\omega N}^2\rho^2\omega^2
\end{eqnarray}
and the pressure
\begin{eqnarray}
P & = & \frac{1}{3\pi^2}\sum_{i=n,p}\int_0^{p_F^i}\frac{p^4}{\sqrt{p^2+M_N^{\ast2}}}dp - \frac{1}{2}m_\sigma^2\sigma^2-\frac{1}{4}g_3\sigma^4 \nonumber \\
& & + \frac{1}{2}m_\omega^2\omega^2+\frac{1}{4}c_3\omega^2 + \frac{1}{2}m_\rho^2\rho^2 \nonumber \\
& & + \frac{1}{2}\Lambda_vg_{\rho N}^2g_{\omega N}^2\rho^2\omega^2
\end{eqnarray}
can be obtained from the Legendre transformation.

We write here expressively also important quantities used for determining our parameters. The incompressibility $K$ at saturation is 
\begin{eqnarray}
K & = & 9\frac{dP(\rho_N,\beta)}{d\rho_N}\bigg|_{\beta = 0,~\rho_N = \rho_0} \nonumber \\
& = & \frac{3p_F^2}{E_F}+\frac{3M_N^\ast p_F}{E_F}\frac{dM_N^\ast}{dp_F} + \frac{9g_{\omega N}^2}{m_\omega^2+3c_3\omega^2}\rho_0,
\end{eqnarray}
and the symmetry energy at saturation is
\begin{eqnarray}
J =\frac{1}{2}\frac{\partial^2E(\rho_N,\beta)}{\partial \beta^2}\bigg|_{\beta = 0,~\rho_N = \rho_0}=\frac{p_F^2}{6E_F}+\frac{g_{\rho N}^2}{2m_\rho^{\ast2}}\rho_0,
\end{eqnarray}
where $\beta = (\rho_n-\rho_p)/\rho_N$ is called neutron-excess parameter, $E(\rho_N,\beta)$ is the binding energy, $p_F = p_F^n = p_F^p$ and $E_F= E_F^n = E_F^p$. 
The slope of symmetry energy $L$ at saturation is defined as
\begin{eqnarray}
L = 3\rho_0\frac{\partial J(\rho_N)}{\partial \rho_N}\bigg|_{\rho_N = \rho_0}.
\end{eqnarray}

\section{Results and Discussions}

\begin{figure}
\centering
\includegraphics[width=8.5cm]{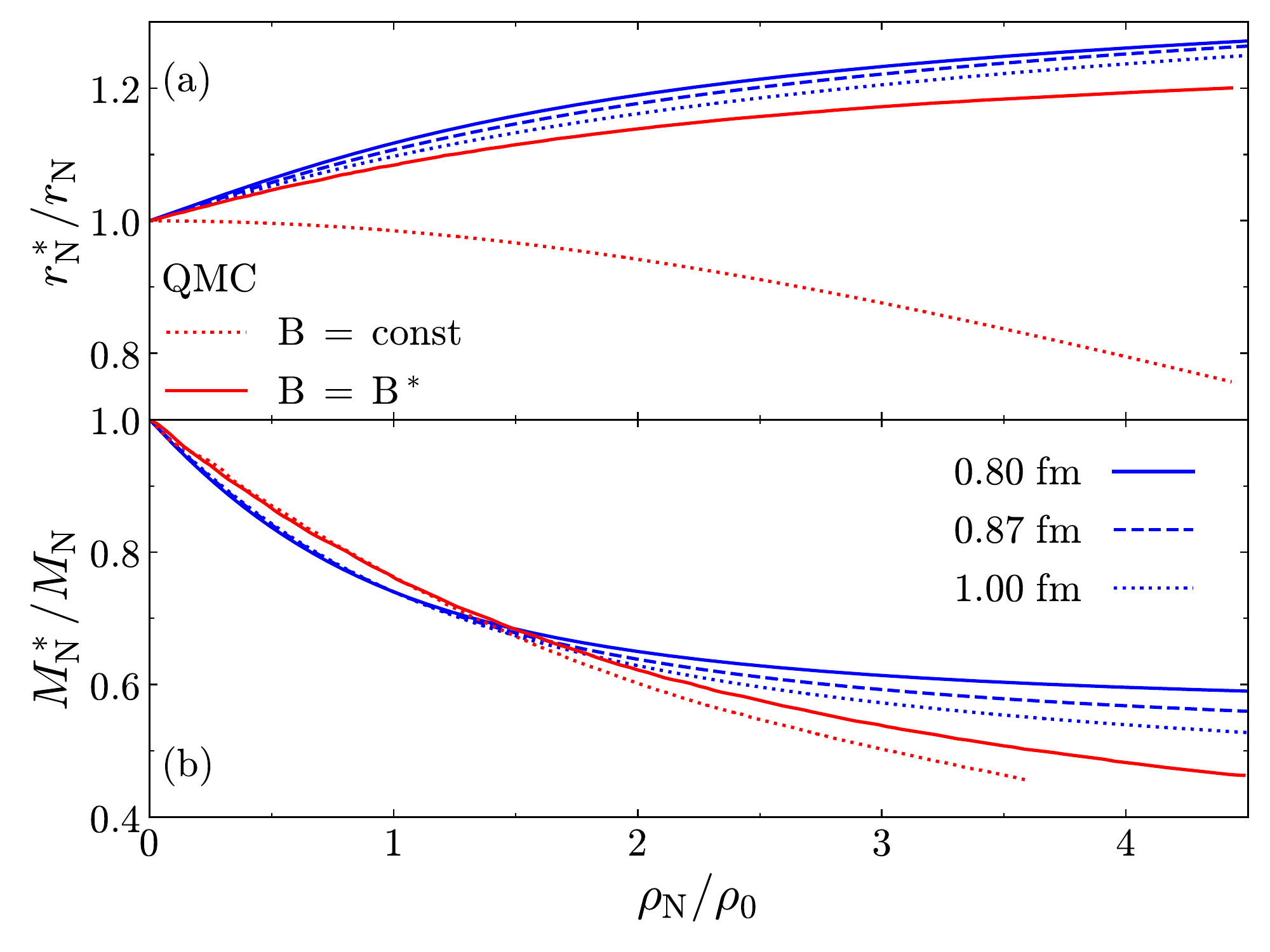}
\caption{(Color online) (Left) effective nucleon radius and (right) effective nucleon mass as a function of $\rho_N/\rho_0$ for symmetric nuclear matter within QMF. Three different free nucleon radius $r_N$ = 0.80 fm, 0.87 fm and 1.00 fm are displayed, with fixed $K = 260$ MeV, $J = 31$ MeV, $L = 60$ MeV. The corresponding QMC results with constant bag (in the case of $r_N = 1.00$ fm)~\cite{radiusform} or density-dependent bag (in the case of $r_N = 0.80$ fm)~\cite{excludevolume} are also shown for comparison.}\label{fig1}
\end{figure}

The effective nucleon radius (mass) as a function of $\rho_N/\rho_0$ for symmetric nuclear matter is shown in the left (right) panel of Fig.~1. The calculations are done with $K = 260$ MeV, $J = 31$ MeV, $L = 60$ MeV, and elsewhere in the following if not specified. The results with three different nucleon radii are displayed: $r_N = 0.80$ fm (solid), $0.87$ fm (dashed) and $1.00$ fm (dotted), to be compared with the QMC results with constant bag~\cite{radiusform} and density-dependent bag~\cite{excludevolume}. We see immediately that the QMF results at all cases accord to the QMC result with density-dependent bag, i.e., $r^*_N$ increases with density, and are opposite to the decreasing behaviour in constant-bag case of QMC. Also, with smaller $r_N$, the increase of $r^*_N$ with density is more pronounced, which can be understood from the excluded volume effects mentioned above. However, the dependence of $M^*_N$ on $r_N$ in QMF is quite different with that in QMC, in the cases of both constant bag~\cite{proton,excludevolume,mbag2} and density-dependent bag~\cite{excludevolume}. In QMC, $M^*_N$ drops with decreasing $r_N$, more phenomenal in the constant-bag case, while in QMF $M^*_N$ increases with decreasing $r_N$.  This advantageously enlarges the range of model applicability for QMF, and should originate from different confining mechanism in the two models.

\begin{figure}
\centering
\includegraphics[width=7.5cm]{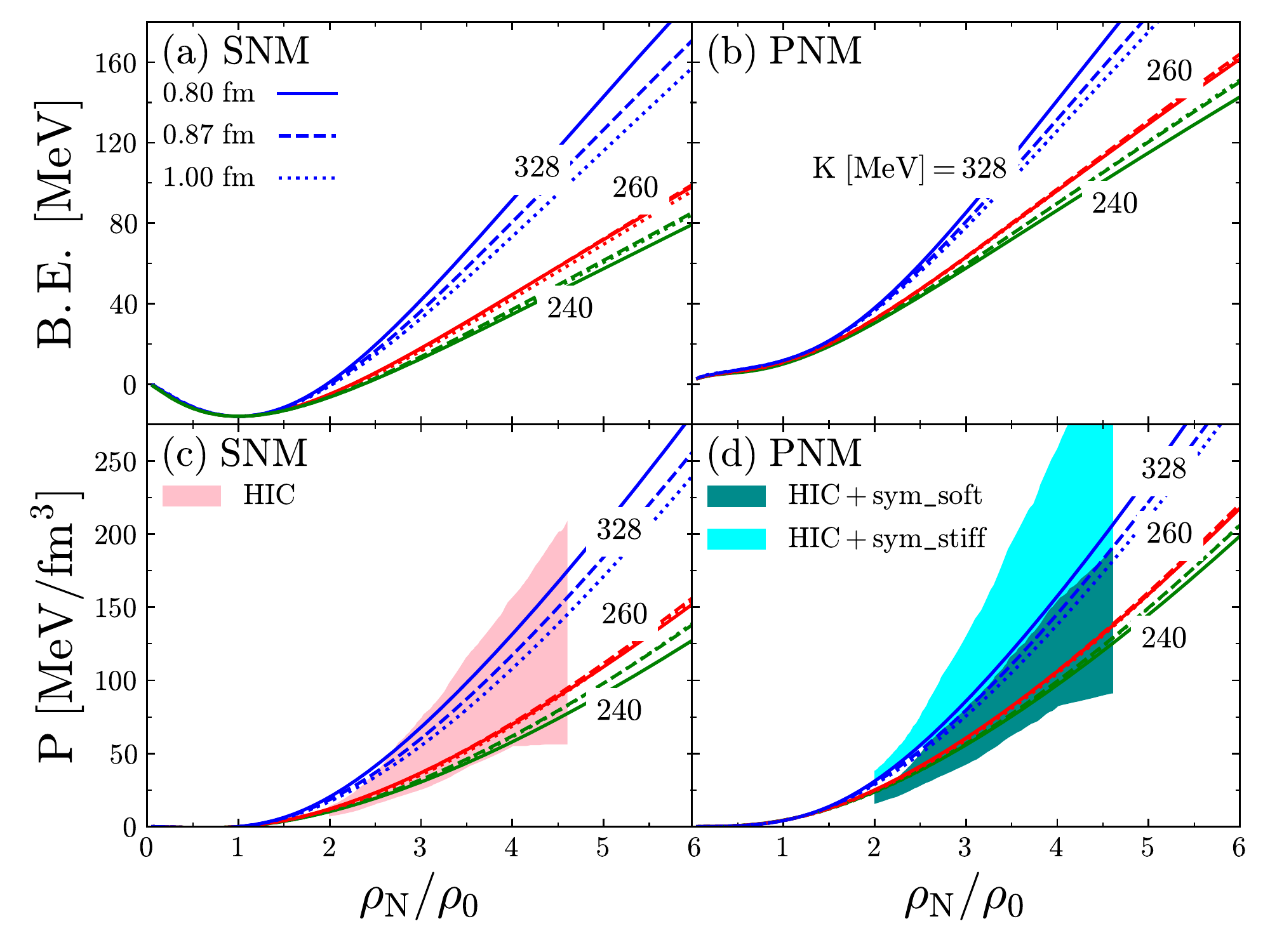}
\caption{(Color online) Binding energy and pressure as a function of $\rho_N/\rho_0$ for symmetric nuclear matter and pure neutron matter. $J = 31$ MeV and $L = 60$ MeV are fixed with $K =$ 240 MeV, 260 MeV, 328 MeV. Results with $r_N = 0.80$ fm, $0.87$ fm, $1.00$ fm are shown in solid, dashed, dotted curves, respectively. The constraints from collective flow in heavy-ion collisions (HIC)~\cite{nmexperiment} are also shown in  the shaded areas, with two density-dependent cases of symmetry energy (light blue for stiff one and dark blue for soft one).}\label{fig2}
\end{figure}

The properties of symmetric nuclear matter (SNM) and pure neutron matter (PNM) are investigated as well. The pressure and binding energy as a function of $\rho_N/\rho_0$ are displayed in Fig.~2, along with the corresponding available experimental constrains (in panel b, d)~\cite{nmexperiment} in shaded areas. We can see that the QMF results in all chosen cases are consistent with the analysis of the elliptic flow from heavy-ion experiments~\cite{nmexperiment} for supranuclear densities above $2\rho_0$. The agreements are better than that in Ref.~\cite{nuclradius} for the interested range of $r_N \sim 0.80$ fm $- 1.00$ fm, since the symmetry energy and its slope are kept same for changing $r_N$ in the present model. The $r_N$ effects are actually small for largest empirical value of $K = 260$ MeV, and only become evident for extreme case of $K = 328$ MeV. 

\begin{figure}
\centering
\includegraphics[width=8.cm]{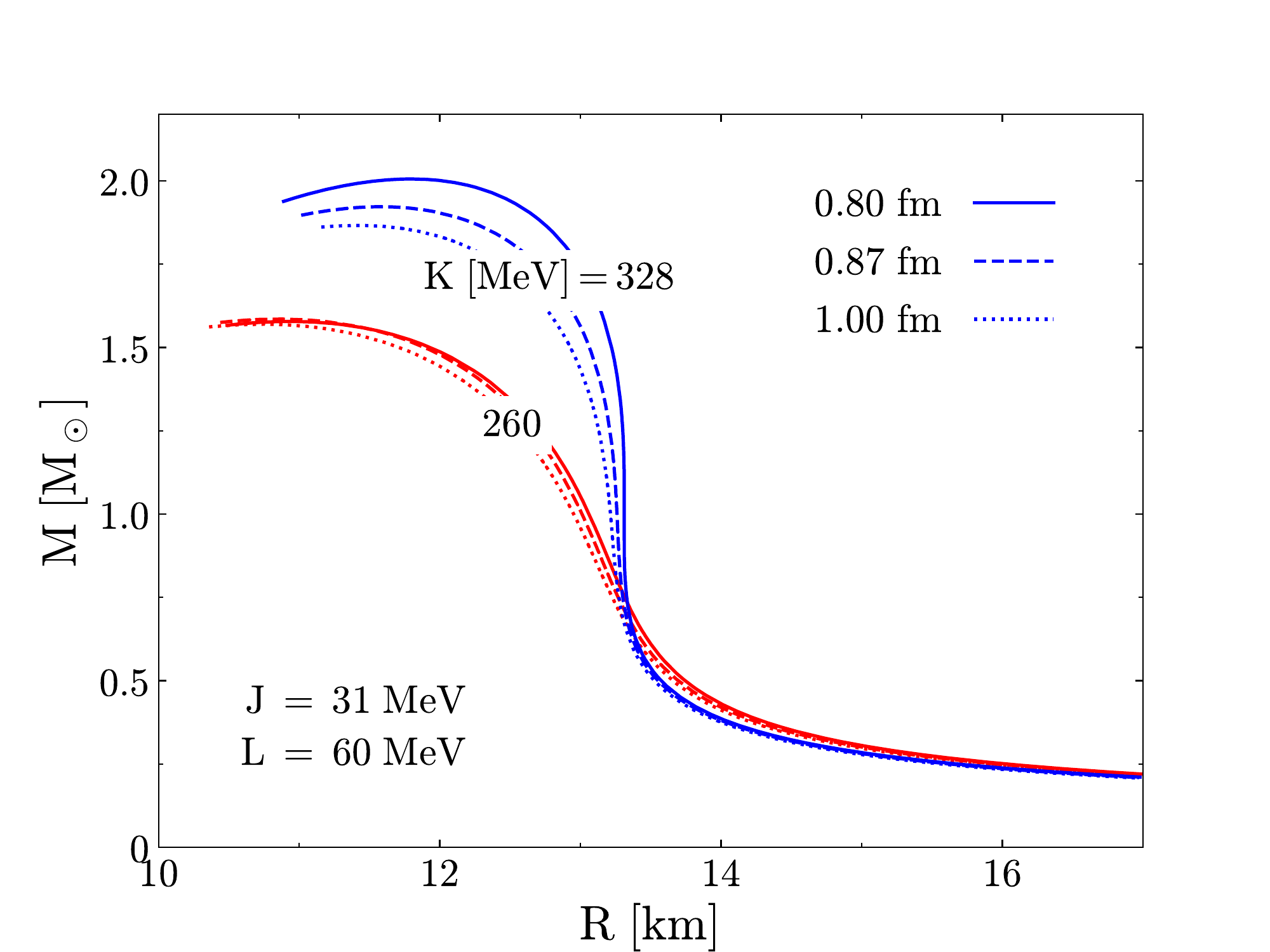}
\caption{(Color online) Mass-radius relations of NSs with $r_N = 0.80$ fm, $0.87$ fm, $1.00$ fm. $J = 31$ MeV and $L = 60$ MeV are fixed with $K = 260$ MeV, 328 MeV.}\label{fig3}
\end{figure}
\begin{figure}
\centering
\includegraphics[width=8.5cm]{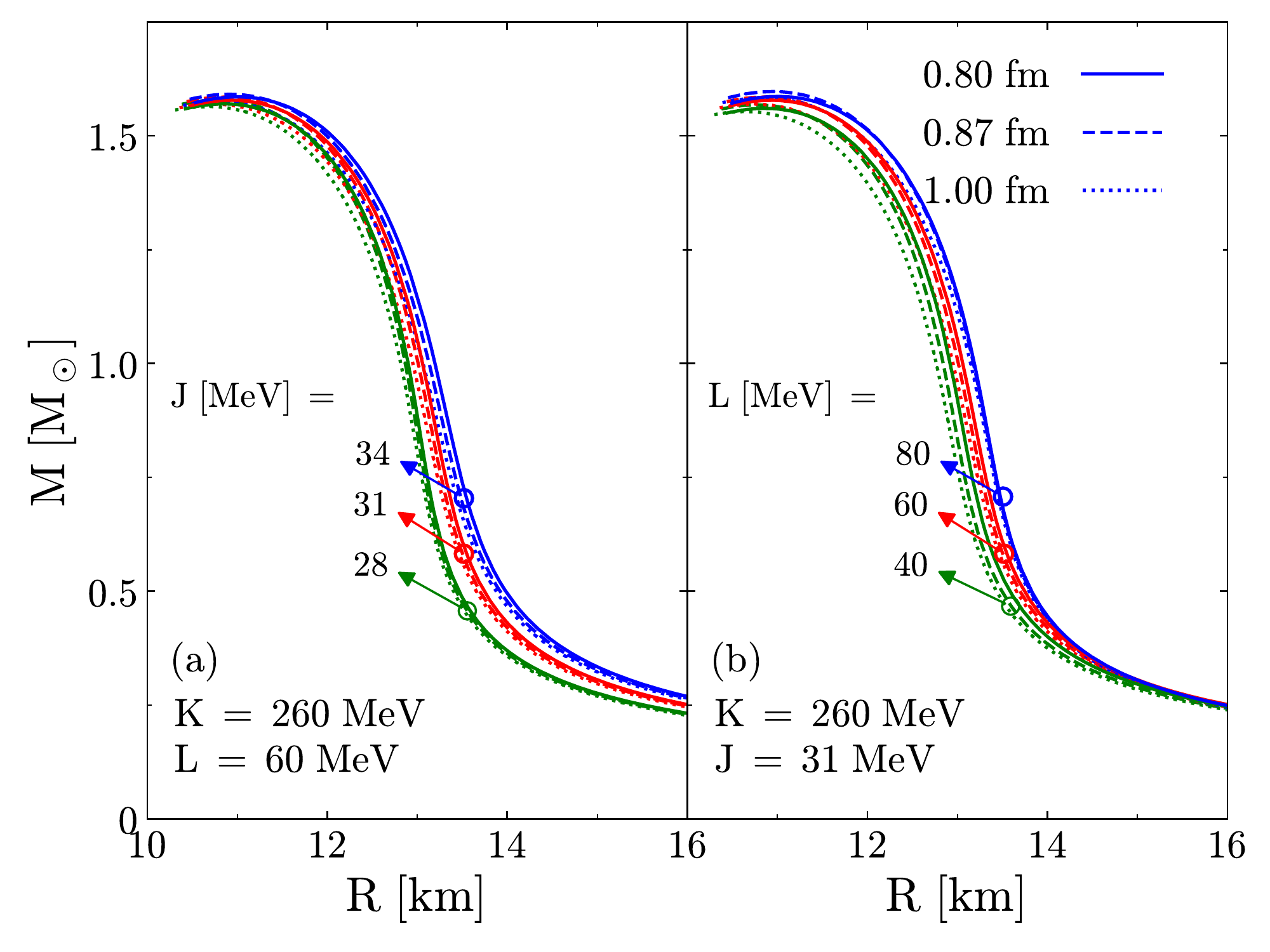}
\caption{(Color online) Same with Fig.~3, but with $K = 260$ MeV fixed, and three values of $J = 28$ MeV, $31$ MeV, $34$ MeV at $L = 60$ MeV (left panel), or three values of $L = 40$ MeV, $60$ MeV, $80$ MeV at $J = 31$ MeV (right panel).}\label{fig4}
\end{figure}

The corresponding mass-radius relations of NSs are presented in Fig.~3, which show clearly that $r_N$ affects little the star radius with empirical saturation properties of $K = 260$ MeV, $J = 31$ MeV and $L = 60$ MeV in our QMF model. If extreme value of $K = 328$ MeV is chosen, decreasing $r_N$ from $1.00$ fm to $0.80$ fm could bring down the maximum mass from $2.01 M_{\odot}$ to $1.87 M_{\odot}$, and the corresponding radii from 11.8 km to 11.4 km (around $3 \%$). This conclusion is different from that in Ref.~\cite{proton,nuclradius}, where they modified $r_N$ without the constrains of fixing symmetry energy (and its slope) at saturation density and found a decreasing star radius $R$ with increasing nucleon radius $r_N$, while the $r_N$-vs-$R$ dependence is very limited in our present study when the symmetry energy impact is excluded. 

To explore further if the change of $R$ is actually from the different symmetry parameters, we present in Fig.~4 the results with fixed $K = 260$ MeV, and modified $J$/$L$ within empirical ranges in the left/right panel. From the left/right panel one can notice that a change of $J$/$L$ in the range of (28 MeV - 34 MeV)/(40 MeV - 80 MeV) could result in a $\sim 1.8 \%$/$\sim 2.9 \%$ variation of $R$ for fixed $r_N$. The variation brought by changing $r_N$ in all cases of fixed ($J,~L$) are even smaller. Therefore, we could conclude that the effect on NS radius $R$ may primarily come from the symmetry energy (i.e., the well-accepted $L$-vs-$R$ dependence), instead of nucleon radius $r_N$. 

\section{Conclusions}

We have studied the effects of free space nucleon radius on NSs by using the QMF model, where nucleons and nuclear matter can be treated self-consistently. The parameters of confinement potential for quarks are obtained by fitting the mass and radius of nucleon in free space, and the nucleon radius are varied by $10 \%$ around the usual $r_N = 0.87$ fm for our propose. The parameters in nucleon-nucleon interaction have also been adjusted so that the properties of symmetric nuclear matter at saturation density satisfy the experimental constrains. 

We have shown the interplay of nucleon radius, incompressibility, symmetry energy and its slope on NS mass-radius relations and found a different conclusion with Ref.~\cite{proton,nuclradius}. The effects of the nucleon radius are weak both on maximum mass and NS radius. Comparatively, the effects of symmetry energy and its slope on NS radius are more obvious. On the other hand, the adjustment of Ref.~\cite{nuclradius} has neglected the symmetry energy fitting. Therefore we argue that the significant influence on NS radius might be from the symmetry energy and its slope, not from the free space nucleon radius. NS radius and the free-space nucleon radius do not have a sensitive dependence. 

For future plans, we notice in Fig.~4 that the NS maximum mass is around 1.6 $M_{\odot}$ and not subject to the uncertainties in $J, L$ within QMF. Although in the extreme $K = 326$ MeV case we could obtain a maximum mass as large as 2.0 $M_{\odot}$ to meet the 2-solar-mass constrain, extra repulsion should be introduced in the model, e.g., by the inclusion of Fock term.

\section{Acknowledgments}
The work was supported by the National Natural Science Foundation of China (No. U1431107).

\end{document}